\documentclass[11pt,aps,prl,superscriptaddress]{revtex4-1}
\usepackage[latin9]{inputenc}
\usepackage{amsmath}

\makeatletter

\providecommand{\tabularnewline}{\\}

\usepackage{amsmath}
\usepackage{graphicx}

\makeatother

\begin{document}

\title{Comment on 'Quantum theory of collective strong coupling of molecular
vibrations with a microcavity mode'}

\author{Luis A. Martínez-Martínez}

\affiliation{Department of Chemistry and Biochemistry, University of California
San Diego, La Jolla, California 92093, United States}

\author{Joel Yuen-Zhou}

\affiliation{Department of Chemistry and Biochemistry, University of California
San Diego, La Jolla, California 92093, United States}

\date{\today}
\begin{abstract}
We have found missing terms and incorrect signs in the secular master
equations reported by Del Pino \emph{et. al.}, \emph{New J. Phys.,
17 (2015), 053040} for vibrational polariton relaxation. Inclusion
of these terms and signs are essential to yield correct (vanishing)
pure dephasing rates between polariton states, as well as coherence
transfer pathways between polaritons and dark states. We provide corrected
expressions for the master equations as well as comparisons with the
results reported by the authors. Even though the main conclusions
of the article are not significantly altered, the corrections are
important to provide a proper description of all possible polariton
relaxation mechanisms within the invoked approximations, especially
when applying the theory to model nonlinear spectroscopy of vibrational
polaritons.
\end{abstract}
\maketitle
\maketitle

The model introduced in \cite{del_pino} by Del Pino and coworkers
(hereafter referred to as DP) consists of an ensemble of $N$ molecular
harmonic vibrational modes with frequency $\omega_{0}$ strongly coupled
to a single microcavity photonic mode at the same frequency (\emph{i.e.},
at resonance with the vibrational modes). The former are in turn coupled
to a rovibrational environment whose spatial extent features two limiting
situations: in one case, the $N$ modes are coupled to a common bath;
in the other case, each vibrational mode interacts with its own independent
but statistically identical bath. Owing to resonance between the vibrational
modes and the photon, the resulting polariton states are $|\pm\rangle=\frac{1}{\sqrt{2}}(a^{\dagger}|0\rangle\pm|B\rangle)$,
where $|B\rangle=\frac{1}{\sqrt{N}}\sum_{n=1}^{N}c_{n}^{\dagger}|0\rangle$
is the totally-symmetry bright state that couples to the microcavity
photon. Assuming periodic boundary conditions, the dark states are
orthogonal to $|B\rangle$, $|d\rangle=\frac{1}{\sqrt{N}}\sum_{n=1}^{N}e^{i\frac{2\pi}{N}dn}c_{n}^{\dagger}|0\rangle$
with nonzero quasimomentum $d=1,\cdots,N-1$. Since only the matter
part of the polaritons interacts with the rovibrational environment,
and both polaritons feature the same matter wavefunction $|B\rangle$
(up to a phase), it follows that both $|+\rangle$ and $|-\rangle$
couple equally to the environment, giving rise to no pure-dephasing
contributions for $\langle\pm|\rho|\mp\rangle$. However, an evaluation
of master equations (12) and (13) in DP yield non-zero pure-dephasing
rates for these coherences (see Table 1 below). This observation led
us to suspect that the aforementioned equations contain mistakes,
which we aim to correct in this note.

To begin with, we point out that the bath correlation functions defined
right before equation (6) in DP are missing some terms,
\begin{align}
\phi_{ij}(t-t') & =\sum_{k}Tr_{b}\left\{ \lambda_{ik}\lambda_{jk}(\tilde{b}_{ik}(t)+\tilde{b}_{ik}^{\dagger}(t))(\tilde{b}_{jk}(t')+\tilde{b}_{jk}^{\dagger}(t'))\right\} .\label{eq:bath correlation}
\end{align}
This corrected definition is important to have consistency with equation
(6) in DP. Next, we rewrite the master equation in the interaction
picture, equation (7) in DP, but given our boundary conditions, we
take care of properly assuming that the eigenstate-site overlap coefficients
$u_{ip}=\langle p|i\rangle$ (where $|i\rangle=c_{i}^{\dagger}|0\rangle$)
can be complex-valued in general,
\begin{equation}
\partial_{t}\tilde{\rho}(t)=\sum_{ij}\sum_{pqrs}\int_{0}^{\infty}u_{ip}u_{qi}u_{jr}u_{sj}e^{i(\omega_{pq}-\omega_{sr})t+i\omega_{sr}\tau}\left[|r\rangle\langle s|\tilde{\rho}(t),|p\rangle\langle q|\right]\phi_{ij}(\tau)d\tau+\text{h.c.}\label{first_redfield}
\end{equation}
Here, $i,j$ and $p,q,r,s$ are site and system-eigenstate indices,
respectively. 

In the secular approximation, only non-oscillatory terms $(\omega_{pq}-\omega_{sr}=0)$
give non-negligible contributions to the dynamics of $\tilde{\rho}$;
this assumption decouples the time evolution for populations and coherences.
The possible combinations $\{p,q,r,s\}$ that satisfy the secular
condition are enumerated in Table 1 in DP; however, they have ommited
$\{p,q,r,s\}$ terms of the form $\{\pm,\pm,\mp,\mp\},\{\pm,\pm,d,d'\},\{d,d',\pm,\pm\}$.
The latter are important to account for the proper evolution of coherences
between polaritons and between polaritons and dark states, as will
be shown next. 

We find that the corrected master equation in the Schrödinger picture
for a common bath ($\phi_{ij}(\tau)=\phi(\tau)$) is given by

\begin{subequations}\label{eq:deloc}

\begin{align}
\partial_{t}\rho & =-i[H_{S},\rho]+\frac{\gamma_{a}}{4}\mathcal{L}_{\sigma_{+-}}[\rho]+\frac{\gamma_{e}}{4}\mathcal{L}_{\sigma_{-+}}[\rho]+\frac{\gamma_{\phi}}{4}\sum_{p=+,-}\mathcal{L}_{\sigma_{pp}}[\rho]+\gamma_{\phi}\mathcal{L}_{\mathcal{D}}[\rho]\label{eq:deloc_a}\\
 & +\frac{\gamma_{\phi}}{4}\left(\sigma_{++}\rho\sigma_{--}^{\dagger}+\sigma_{--}\rho\sigma_{++}^{\dagger}\right)+\frac{\gamma_{\phi}}{2}\sum_{d}\left(\sigma_{++}\rho\sigma_{dd}^{\dagger}+\sigma_{dd}\rho\sigma_{++}^{\dagger}\right)\label{eq:deloc_b}\\
 & +\frac{\gamma_{\phi}}{2}\sum_{d}\left(\sigma_{--}\rho\sigma_{dd}^{\dagger}+\sigma_{dd}\rho\sigma_{--}^{\dagger}\right),\label{eq:deloc_b_bis}
\end{align}

\end{subequations}\noindent where $\sigma_{ab}=|a\rangle\langle b|$.
Here, the terms in equation (\ref{eq:deloc_a}) are identical to those
in equation (12) in DP; however, the authors have missed the terms
in equations (\ref{eq:deloc_b}) and (\ref{eq:deloc_b_bis}), which
arise from the omitted secular contributions outlined above. 

On the other hand the master equation corresponding to localized baths
($\phi_{ij}(\tau)=\delta_{ij}\phi(\tau)$) reads

\begin{subequations}\label{eq:loc}

\begin{align}
\partial_{t}\rho & =-i[H_{s},\rho]+\frac{\gamma_{a}}{4N}\mathcal{L}_{\sigma_{+-}}[\rho]+\frac{\gamma_{e}}{4N}\mathcal{L}_{\sigma_{-+}}[\rho]\label{eq:correct1}\\
 & +\sum_{d}\frac{\Gamma_{a}}{2N}(\mathcal{L}_{\sigma_{d-}}[\rho]+\mathcal{L}_{\sigma_{+d}}[\rho])+\sum_{d}\frac{\Gamma_{e}}{2N}(\mathcal{L}_{\sigma_{d+}}[\rho]+\mathcal{L}_{\sigma_{-d}}[\rho])\label{eq:correct2}\\
 & +\frac{\Gamma_{a}}{4N}\sum_{d}\left(-\left[\sigma_{\bar{d}-}\rho,\sigma_{d+}\right]-\left[\sigma_{+\bar{d}}\rho,\sigma_{-d}\right]+\text{h.c.}\right)\label{eq:error1}\\
 & +\frac{\Gamma_{e}}{4N}\sum_{d}\left(-\left[\sigma_{\bar{d}+}\rho,\sigma_{d-}\right]-\left[\sigma_{-\bar{d}}\rho,\sigma_{+d}\right]+\text{h.c.}\right)\label{eq:error2}\\
 & +\frac{\gamma_{\phi}}{4N}\sum_{p=+,-}\mathcal{L}_{\sigma_{pp}}[\rho]+\gamma_{\phi}\sum_{i}\mathcal{L}_{\mathcal{D}c_{i}^{\dagger}c_{i}\mathcal{D}}[\rho]\label{eq:correct3}\\
 & +\frac{\gamma_{\phi}}{4N}\left(\sigma_{++}\rho\sigma_{--}^{\dagger}+\sigma_{--}\rho\sigma_{++}^{\dagger}\right)+\frac{\gamma_{\phi}}{2N}\sum_{d}\left(\sigma_{++}\rho\sigma_{dd}^{\dagger}+\sigma_{dd}\rho\sigma_{++}^{\dagger}\right)\label{eq:missing1}\\
 & +\frac{\gamma_{\phi}}{2N}\sum_{d}\left(\sigma_{--}\rho\sigma_{dd}^{\dagger}+\sigma_{dd}\rho\sigma_{--}^{\dagger}\right)\label{eq:missing2}
\end{align}

\end{subequations} Here, to avoid confusion, we use the notation
$|\bar{d}\rangle\equiv|N-d\rangle$ (the dark state with opposite
quasimomentum to that of $|d\rangle$). The terms in equations (\ref{eq:correct1}),
(\ref{eq:correct2}), and (\ref{eq:correct3}) coincide with those
in equations (13a), (13b), and (13e) in DP, respectively. Equations
(\ref{eq:error1}) and (\ref{eq:error2}) differ from those in equations
(13c) and (13d) in DP by some signs. Finally, equations (\ref{eq:missing1})
and (\ref{eq:missing2}) are missing in DP and arise from the omitted
secular terms.

For a better appreciation of the missing information in DP, we compare
in Table 1 the equations of motion for populations and coherences
that follow from Eqs. (\ref{eq:deloc}) and (\ref{eq:loc}) with those
that follow from DP. By reporting the results in the interaction picture,
we simply neglect the trivial coherent dynamics generated by $H_{s}$.

\begin{table}
\begin{tabular}{|c|c|c|}
\hline 
$\partial_{t}\tilde{\rho}_{ab}$ & Master equations in \cite{del_pino} & Corrected master equations\tabularnewline
\hline 
\hline 
\multicolumn{3}{|c|}{Delocalized bath case}\tabularnewline
\hline 
$\partial_{t}\tilde{\rho}_{+-}$ & $\left(-\frac{\gamma_{\phi}}{4}-\frac{\gamma_{a}}{8}-\frac{\gamma_{e}}{8}\right)\tilde{\rho}_{+-}$ & $\underbrace{\left(-\frac{\gamma_{e}}{8}-\frac{\gamma_{a}}{8}\right)\tilde{\rho}_{+-}}_{T_{1}\,\text{term}}$\tabularnewline
\hline 
$\partial_{t}\tilde{\rho}_{+d}$ & $\left(-\frac{\gamma_{e}}{8}-\frac{5\gamma_{\phi}}{8}\right)\tilde{\rho}_{+d}$ & $\begin{array}{c}
\underbrace{-\frac{\gamma_{e}}{8}\tilde{\rho}_{+d}}_{T_{1\,}\text{term}}\\
\underbrace{-\frac{\gamma_{\phi}}{8}\tilde{\rho}_{+d}}_{T_{2}^{*}\,\text{term}}
\end{array}$\tabularnewline
\hline 
$\partial_{t}\tilde{\rho}_{d_{1}d_{2}}$ & 0 & 0\tabularnewline
\hline 
$\partial_{t}\tilde{\rho}_{++}$ & $-\frac{\gamma_{e}}{4}\tilde{\rho}_{++}+\frac{\gamma_{a}}{4}\tilde{\rho}_{--}$ & $-\frac{\gamma_{e}}{4}\tilde{\rho}_{++}+\frac{\gamma_{a}}{4}\tilde{\rho}_{--}$\tabularnewline
\hline 
$\partial_{t}\tilde{\rho}_{dd}$ & 0 & 0\tabularnewline
\hline 
\multicolumn{3}{|c|}{Localized bath case}\tabularnewline
\hline 
$\partial_{t}\tilde{\rho}_{+-}$ & $-\left(\frac{\gamma_{a}}{8N}+\frac{\gamma_{e}}{8N}+\frac{\Gamma_{a}}{4N}(N-1)+\frac{\Gamma_{e}}{4N}(N-1)+\frac{\gamma_{\phi}}{8N}\right)\tilde{\rho}_{+-}$ & $\underbrace{-\left(\frac{\gamma_{e}}{8N}+\frac{\Gamma_{e}}{4N}(N-1)+\frac{\gamma_{a}}{8N}+\frac{\Gamma_{a}}{4N}(N-1)\right)\tilde{\rho}_{+-}}_{T_{1}\,\text{term}}$\tabularnewline
\hline 
$\partial_{t}\tilde{\rho}_{+d}$ & $\begin{array}{c}
-\big(\frac{\gamma_{e}}{8N}+\frac{\Gamma_{a}}{4N}+\frac{\Gamma_{e}}{4N}(N-1)\big)\tilde{\rho}_{+d}\\
-\big(\frac{\Gamma_{e}}{4N}+\frac{\gamma_{\phi}}{8N}+\frac{\gamma_{\phi}}{2N}(N-1)\big)\tilde{\rho}_{+d}
\end{array}$ & $\begin{array}{c}
\underbrace{-\big(\frac{\gamma_{e}}{8N}+\frac{\Gamma_{e}}{4N}(N-1)+\frac{\Gamma_{a}}{4N}+\frac{\Gamma_{e}}{4N}+\frac{\gamma_{\phi}}{2N}(N-2)\big)\tilde{\rho}_{+d}}_{T_{1}\,\text{term}}\\
\underbrace{-\frac{\gamma_{\phi}}{8N}\tilde{\rho}_{+d}}_{T_{2}^{*}\,\text{term}}\\
\underbrace{-\frac{\Gamma_{a}}{2N}\tilde{\rho}_{\bar{d}-}}_{\text{coherence transfer}}
\end{array}$\tabularnewline
\hline 
$\partial_{t}\tilde{\rho}_{d_{1}d_{2}}$ & $\begin{array}{c}
-\Big(\frac{\Gamma_{a}}{2N}+\frac{\Gamma_{e}}{2N}+\frac{\gamma_{\phi}}{N}(N-2)\Big)\tilde{\rho}_{d_{1}d_{2}}\\
+\frac{\gamma_{\phi}}{N}\sum_{d,d'}(\delta_{d_{1}-d_{2},d-d'}-\delta_{d,d_{1}}\delta_{d',d_{2}})\tilde{\rho}_{dd'}
\end{array}$ & $\begin{array}{c}
\underbrace{-\Big(\frac{\Gamma_{a}}{2N}+\frac{\Gamma_{e}}{2N}+\frac{\gamma_{\phi}}{N}(N-2)\Big)\tilde{\rho}_{d_{1}d_{2}}}_{T_{1}\,\text{term}}\\
\underbrace{+\frac{\gamma_{\phi}}{N}\sum_{d,d'}(\delta_{d_{1}-d_{2},d-d'}-\delta_{d,d_{1}}\delta_{d',d_{2}})\tilde{\rho}_{dd'}}_{\text{coherence transfer}}
\end{array}$\tabularnewline
\hline 
$\partial_{t}\tilde{\rho}_{++}$ & $\begin{array}{c}
-\frac{\gamma_{e}}{4N}\tilde{\rho}_{++}+\frac{\gamma_{a}}{4N}\tilde{\rho}_{--}-\frac{\Gamma_{e}}{2N}(N-1)\tilde{\rho}_{++}\\
+\sum_{d}\frac{\Gamma_{a}}{2N}\tilde{\rho}_{dd}
\end{array}$ & $\begin{array}{c}
-\frac{\gamma_{e}}{4N}\tilde{\rho}_{++}+\frac{\gamma_{a}}{4N}\tilde{\rho}_{--}-\frac{\Gamma_{e}}{2N}(N-1)\tilde{\rho}_{++}\\
+\sum_{d}\frac{\Gamma_{a}}{2N}\tilde{\rho}_{dd}
\end{array}$\tabularnewline
\hline 
$\partial_{t}\tilde{\rho}_{dd}$ & $\begin{array}{c}
\frac{\Gamma_{e}}{2N}\tilde{\rho}_{++}-\frac{\Gamma_{a}}{2N}\tilde{\rho}_{dd}+\frac{\Gamma_{a}}{2N}\tilde{\rho}_{--}\\
-\frac{\Gamma_{e}}{2N}\tilde{\rho}_{dd}-\frac{\gamma_{\phi}}{N}(N-2)\tilde{\rho}_{dd}+\frac{\gamma_{\phi}}{N}\sum_{d'\neq d}\tilde{\rho}_{d'd'}
\end{array}$ & $\begin{array}{c}
\frac{\Gamma_{e}}{2N}\tilde{\rho}_{++}-\frac{\Gamma_{a}}{2N}\tilde{\rho}_{dd}+\frac{\Gamma_{a}}{2N}\tilde{\rho}_{--}\\
-\frac{\Gamma_{e}}{2N}\tilde{\rho}_{dd}-\frac{\gamma_{\phi}}{N}(N-2)\tilde{\rho}_{dd}+\frac{\gamma_{\phi}}{N}\sum_{d'\neq d}\tilde{\rho}_{d'd'}
\end{array}$\tabularnewline
\hline 
\end{tabular}

\caption{Comparison of equations of motion in the interaction picture for coherences
$\left(\langle a|\tilde{\rho}|b\rangle=\tilde{\rho}_{ab},\,a\protect\neq b\right)$
and populations $\left(\langle a|\tilde{\rho}|a\rangle=\tilde{\rho}_{aa}\right)$
of the reduced vibrational-polariton density matrix calculated by
DP and using the corrected equations (\ref{eq:deloc}) and (\ref{eq:loc}).
Additional expressions can be obtained by simultaneously making the
changes $+\leftrightarrow-$ and $\Gamma_{a}\leftrightarrow\Gamma_{e}$
throughout (for instance, we can obtain $\partial_{t}\tilde{\rho}_{-+}$
and $\partial_{t}\tilde{\rho}_{-d}$ from $\partial_{t}\tilde{\rho}_{-+}$
and $\partial_{t}\tilde{\rho}_{+d}$, respectively). Also, $\partial_{t}\tilde{\rho}_{\pm d}=\partial_{t}\tilde{\rho}_{d\pm}^{*}$
and $d$ labels dark states. Here, the $\delta_{d_{1}-d_{2},d-d'}$
term indicates conservation of vibrational quasimomentum and must
be interpreted in terms of mod $N$ arithmetic.}
\end{table}
We now briefly elaborate on the physical processes that are incompletely
captured by DP. First, in secular Redfield theory, the decay rate
$T_{2}^{-1}$ of a coherence $\rho_{ab}$ has several contributions:
one is an average of the population decay rates of the system states
$|a\rangle$ and $|b\rangle$, often labeled $T_{1}^{-1}$; the other
one is the pure dephasing rate $T_{2}^{*-1}$ associated with fluctuations
of the system-environment coupling between $|a\rangle$ and $|b\rangle$;
finally, there can also be coherence transfers to $\rho_{cd}$ as
long as $\omega_{cd}=\omega_{ab}$. These physical processes have
been highlighted in Table 1, and can be readily calculated using textbook
formalism such as that found in \cite{may-kuhn}.

We note that some of the omitted terms by DP are proportional to $\gamma_{\phi}$,
indicating pure-dephasing contributions, in agreement with our original
observation that the pure-dephasing rate between $|\pm\rangle$ and
$|\mp\rangle$ is zero; see entries corresponding to $\partial\tilde{\rho}_{+-}$,
which do not feature $T_{2}^{*}$ terms. We also notice that equations
(\ref{eq:error1}) and (\ref{eq:error2}) yield non-zero contributions
for the decay rate of coherences between polariton and dark states
in the independent baths case. This is in contrast with equations
(13c) and (13d) in DP, which vanish identically due to sign errors.
For instance, from Table 1, the corrected coherence evolution $\partial_{t}\rho_{+d}$
contains an additional decay due to coherence transfer $-\frac{\Gamma_{a}}{2N}\langle\bar{d}|\rho|-\rangle$.
This term contributes to an additional decay channel of $\rho_{+d}$
and arises due to the fact that both $|+\rangle$ and $|-\rangle$
feature the same molecular state $|B\rangle$, but with opposite signs. 

Table 1 shows that our corrections do not change the main conclusions
established by DP, since their discussion was mainly focused on population
transfer dynamics, while the omissions affect coherences. These omissions,
however, will be essential to understand nonlinear spectroscopic signals
of polaritons.

\section{Appendix}

\subsection{Missing terms for pure dephasing}

We begin by invoking the secular approximation in equation (\ref{first_redfield}),
setting $\omega_{pq}-\omega_{sr}=0$. We only consider the missing
terms in DP, namely, the cases where $\{p,q,r,s\}$ are $\{\pm,\pm,\mp,\mp\}$,
$\{\pm,\pm,d,d'\}$, and $\{d,d',\pm,\pm\}$. The system-eigenstate-site
overlaps $\left|u_{\pm i}\right|=\frac{1}{\sqrt{2N}}$ and $|u_{di}|=\frac{1}{\sqrt{N}}$
become handy. 

To proceed, let us analyze the two bath cases:
\begin{enumerate}
\item For the common bath, $\phi_{ij}(\tau)=\phi(\tau)$, so $\sum_{ij}u_{ip}u_{qi}u_{jr}u_{sj}\phi(\tau)=\langle p|\mathbf{P}_{vib}|q\rangle\langle r|\mathbf{P}_{vib}|s\rangle\phi(\tau)$,
where $\mathbf{P}_{vib}=\sum_{i}|i\rangle\langle i|$ is the projector
on the vibrational subspace. For the cases of interest above, we have
two possibilities:
\begin{enumerate}
\item $\{p,q,r,s\}=\{\pm,\pm,\mp,\mp\}$, in which case $\langle p|\mathbf{P}_{vib}|q\rangle\langle r|\mathbf{P}_{vib}|s\rangle=\frac{1}{4}$.
\item $\{p,q,r,s\}=\{\pm,\pm,d,d'\},\{d,d',\pm,\pm\}$, in which case $\langle p|\mathbf{P}_{vib}|q\rangle\langle r|\mathbf{P}_{vib}|s\rangle=\frac{\delta_{dd'}}{2}$.
\end{enumerate}
\item For independent baths, $\phi_{ij}(\tau)=\delta_{ij}\phi(\tau)$, so
$\sum_{ij}u_{ip}u_{qi}u_{jr}u_{sj}\delta_{ij}\phi(\tau)=\sum_{i}\langle p|i\rangle\langle i|q\rangle\langle r|i\rangle\langle i|s\rangle\phi(\tau)$.
We analyze the two possibilities again:
\begin{enumerate}
\item $\{p,q,r,s\}=\{\pm,\pm,\mp,\mp\}$, in which case $\sum_{i}\langle p|i\rangle\langle i|q\rangle\langle r|i\rangle\langle i|s\rangle=N\frac{1}{4N^{2}}=\frac{1}{4N}$.
\item $\{p,q,r,s\}=\{\pm,\pm,d,d'\},\{d,d',\pm,\pm\}$, in which case $\sum_{i}\langle p|i\rangle\langle i|q\rangle\langle r|i\rangle\langle i|s\rangle=N\frac{1}{2N^{2}}\delta_{dd'}=\frac{\delta_{dd'}}{2N}$. 
\end{enumerate}
\end{enumerate}
This exercise allows us to discard the $d\neq d'$ cases. Hence, we
only need to develop the $\{p,q,r,s\}=\{p,p,r,r\}$ term (where $p\neq r$)
in the right-hand-side of equation (\ref{first_redfield}),

\begin{equation}
\sum_{ij}\int_{0}^{\infty}u_{ip}u_{pi}u_{jr}u_{rj}\left[|r\rangle\langle r|\tilde{\rho}(t),|p\rangle\langle p|\right]\phi_{ij}(\tau)d\tau+\text{h.c.}=\gamma_{r,p}^{deph}\sigma_{rr}\rho\sigma_{pp}^{\dagger}+\text{h.c.},\label{eq:pprr}
\end{equation}
where

\begin{equation}
\gamma_{r,p}^{deph}=\gamma_{p,r}^{deph}=\sum_{ij}u_{ip}u_{pi}u_{jr}u_{rj}\int_{-\infty}^{\infty}\phi_{ij}(\tau)d\tau.\label{gamma_dep}
\end{equation}
Inclusion of terms of the form of equation of (\ref{eq:pprr}) for
$\{p,r\}=\{\pm,\mp\},\{\pm,d\},\{d,\mp\}$ into the master equation
gives rise to the correction terms in equations (\ref{eq:deloc_b}),
(\ref{eq:deloc_b_bis}), (\ref{eq:missing1}), and (\ref{eq:missing2}),
where $\gamma_{\mp,\pm}^{deph}=\frac{\gamma_{\phi}}{4}$, $\gamma_{d,\pm}^{deph}=\gamma_{\pm,d}^{deph}=\frac{\gamma_{\phi}}{2}$
for the common bath and $\gamma_{\mp,\pm}^{deph}=\frac{\gamma_{\phi}}{4N}$,
$\gamma_{d,\pm}^{deph}=\gamma_{\pm,d}^{deph}=\frac{\gamma_{\phi}}{2N}$
for independent baths, where $\gamma_{\phi}=2S(0)$.

\subsection{Missing coherence transfer pathways}

DP miss coherence transfer pathways that arise from combinations $\left\{ p,q,r,s\right\} =\left\{ +d_{1},-d_{2}\right\} $
in equation (\ref{first_redfield}),

\begin{align}
 & \sum_{ij}\int_{0}^{\infty}u_{i+}u_{d_{1}i}u_{j-}u_{d_{2}j}e^{i\omega_{d-}\tau}\phi_{ij}(\tau)d\tau\left[|-\rangle\langle d_{2}|\tilde{\rho},|+\rangle\langle d_{1}|\right]+\text{h.c.}\nonumber \\
+ & \sum_{ij}\int_{0}^{\infty}u_{id_{1}}u_{-i}u_{jd_{2}}u_{+j}e^{i\omega_{+d}\tau}\phi_{ij}(\tau)d\tau\left[|d_{2}\rangle\langle+|\tilde{\rho},|d_{1}\rangle\langle-|\right]+\text{h.c.}\nonumber \\
+ & \sum_{ij}\int_{0}^{\infty}u_{id_{1}}u_{+i}u_{jd_{2}}u_{-j}e^{i\omega_{-d}\tau}\phi_{ij}(\tau)d\tau\left[|d_{2}\rangle\langle-|\tilde{\rho},|d_{1}\rangle\langle+|\right]+\text{h.c.}\nonumber \\
+ & \sum_{ij}\int_{0}^{\infty}u_{i-}u_{d_{1}i}u_{j+}u_{d_{2}j}e^{i\omega_{d+}\tau}\phi_{ij}(\tau)d\tau\left[|+\rangle\langle d_{2}|\tilde{\rho},|-\rangle\langle d_{1}|\right]+\text{h.c.}\label{eq:coherence_transfer}
\end{align}
where $\omega_{d}=\omega_{d_{1}}=\omega_{d_{2}}$. Notice that due
to orthogonality $\langle\pm|d_{1(2)}\rangle=0$, each of these four
terms is zero for the case of the common bath. For the case of the
independent baths, we use the fact that

\begin{equation}
\sum_{i}u_{i\pm}u_{d_{1}i}u_{i\mp}u_{d_{2}i}=-\frac{1}{2N}\sum_{i}u_{d_{1}i}u_{d_{2}i}=-\frac{1}{2N}\delta_{d_{2},N-d_{1}},\label{eq:momentum conservation}
\end{equation}
where conservation of quasimomentum in the finite $N$ molecule chain
leads to $d_{2}+d_{1}=0(\text{mod}\,N)$, or $d_{2}=N-d_{1}\equiv\bar{d}_{1}$.
The expression in equation (\ref{eq:coherence_transfer}) then becomes,

\begin{align}
 & -\Bigg(\frac{\Gamma_{a}}{4N}\left[|\bar{d}\rangle\langle-|\tilde{\rho},|d\rangle\langle+|\right]+\frac{\Gamma_{a}}{4N}\left[|+\rangle\langle\bar{d}|\tilde{\rho},|-\rangle\langle d|\right]\nonumber \\
 & +\frac{\Gamma_{e}}{4N}\left[|-\rangle\langle\bar{d}|\tilde{\rho},|+\rangle\langle d|\right]+\frac{\Gamma_{e}}{4N}\left[|\bar{d}\rangle\langle+|\tilde{\rho},|d\rangle\langle-|\right]\Bigg)+\text{h.c.}\label{eq:missing_term_coherence_transfer}
\end{align}
where, as usual, we have taken only the real part of the resulting
half-sided Fourier transforms (assuming that the Lamb shift corresponding
to the imaginary part can be absorbed into the coherent dynamics),
$\Re\int_{0}^{\infty}e^{i\omega_{qr}\tau}\phi(\tau)d\tau=S(\omega_{qr})$
and $\Gamma_{a}=2S(\omega_{d+})=2S(\omega_{-d})$ and $\Gamma_{e}=2S(\omega_{+d})=2S(\omega_{d-})$.
Equation (\ref{eq:missing_term_coherence_transfer}) is equal to equations
(\ref{eq:error1}) and (\ref{eq:error2}), which were featured with
incorrect signs as equations (13c) and (13d) in DP.

\bibliographystyle{naturemag}
\bibliography{bibliography_nature_convention}

\end{document}